\documentclass{article}
\usepackage{spconf,amsmath,graphicx}
\usepackage{cite,mathtools,amssymb,bm}
\usepackage{balance}
\usepackage[noend]{algpseudocode}
\usepackage{algorithm}
\usepackage{booktabs}
\usepackage{siunitx}
\usepackage{color}
\title{Streaming automatic speech recognition with the transformer model}

\name{Niko Moritz, Takaaki Hori, Jonathan Le Roux}
\address{Mitsubishi Electric Research Laboratories (MERL), Cambridge, MA, USA}

\begin{document}
\ninept
\setlength{\parindent}{0pt}
\setlength{\abovedisplayskip}{4pt}
\setlength{\belowdisplayskip}{4pt}
\setlength{\textfloatsep}{5pt}

\maketitle
\begin{abstract}

Encoder-decoder based sequence-to-sequence models have demonstrated state-of-the-art results in end-to-end automatic speech recognition (ASR). Recently, the transformer architecture, which uses self-attention to model temporal context information, has been shown to achieve significantly lower word error rates (WERs) compared to recurrent neural network (RNN) based system architectures. Despite its success, the practical usage is limited to offline ASR tasks, since encoder-decoder architectures typically require an entire speech utterance as input. In this work, we propose a transformer based end-to-end ASR system for streaming ASR, where an output must be generated shortly after each spoken word. To achieve this, we apply time-restricted self-attention for the encoder and triggered attention for the encoder-decoder attention mechanism. Our proposed streaming transformer architecture achieves 2.8\% and 7.2\% WER for the ``clean'' and ``other'' test data of LibriSpeech, which to our knowledge is the best published streaming end-to-end ASR result for this task.

\end{abstract}
\begin{keywords}
automatic speech recognition, streaming, end-to-end, transformer, triggered attention
\end{keywords}
\section{Introduction}
\label{sec:intro}

Hybrid hidden Markov model (HMM) based automatic speech recognition (ASR) systems have provided state-of-the-art results for the last few decades \cite{PoveyPGG16,HintonDYD12}. End-to-end ASR systems, which approach the speech-to-text conversion problem using a single sequence-to-sequence model, have recently demonstrated competitive performance \cite{KaritaYWD19}. The most popular and successful end-to-end ASR approaches are based on connectionist temporal classification (CTC) \cite{GravesFGS06}, recurrent neural network (RNN) transducer (RNN-T) \cite{Graves12}, and attention-based encoder-decoder architectures \cite{BahdanauCB14}. RNN-T based ASR systems achieve state-of-the-art ASR performance for streaming/online applications and are successfully deployed in production systems \cite{Schalkwyk19,LiZHG19}. Attention-based encoder-decoder architectures, however, are the best performing end-to-end ASR systems \cite{PrabhavalkarRSL17}, but they cannot be easily applied in a streaming fashion, which prevents them from being used more widely in practice. To overcome this limitation, different methods for streaming ASR with attention-based systems haven been proposed such as the neural transducer (NT) \cite{SainathCPKWNC17}, monotonic chunkwise attention (MoChA) \cite{ChiuR18}, and triggered attention (TA) \cite{MoritzHR19}. The NT relies on traditional block processing with fixed window size and stride to produce incremental attention model outputs. The MoChA approach uses an extra layer to compute a selection probability that defines the length of the output label sequence and provides an alignment to chunk the encoder state sequence prior to soft attention. The TA system requires that the attention-based encoder-decoder model is trained jointly with a CTC objective function, which has also been shown to improve attention-based systems \cite{WatanabeHKHH17}, and the CTC output is used to predict an alignment that triggers the attention decoding process \cite{MoritzHR19}. A frame-synchronous one-pass decoding algorithm for joint CTC-attention scoring was proposed in \cite{MoritzHR19c} to further optimize and enhance ASR decoding using the TA concept.

Besides the end-to-end ASR modeling approach, the underlying neural network architecture is of paramount importance as well to achieve good ASR performance. RNN-based architectures, such as the long short-term memory (LSTM) neural network, are often applied for end-to-end ASR systems. Bidirectional LSTMs (BLSTMs) achieve state-of-the-art results among such RNN-based systems but are unsuitable for application in a streaming fashion, where unidirectional LSTMs or latency-controlled BLSTMs (LC-BLSTMs) must be applied instead \cite{MoritzHR19b}. The parallel time-delayed LSTM (PTDLSTM) architecture has been proposed to further reduce the word error rate (WER) gap between unidirectional and bidirectional architectures and to improve the computational complexity compared to the LC-BLSTM \cite{MoritzHR19b}. Recently, the transformer model, which is an encoder-decoder type of architecture based on self-attention originally proposed for machine translation \cite{VaswaniSPUJGKP17}, has been applied to ASR with promising results and improved WERs compared to RNN-based architectures \cite{KaritaCH19}.

In this work, we apply time-restricted self-attention to the encoder, and the TA concept to the encoder-decoder attention mechanism of the transformer model to enable the application of online/streaming ASR. The transformer model is jointly trained with a CTC objective to optimize training and decoding results as well as to enable the TA concept \cite{KaritaYWD19,MoritzHR19}. For joint CTC-transformer decoding and scoring, we employ the frame-synchronous one-pass decoding algorithm proposed in \cite{MoritzHR19c}.

\section{Streaming Transformer}
\label{sec:trans_architecture}

\begin{figure}[tb]
  \centering
  \centerline{\includegraphics[width=\linewidth]{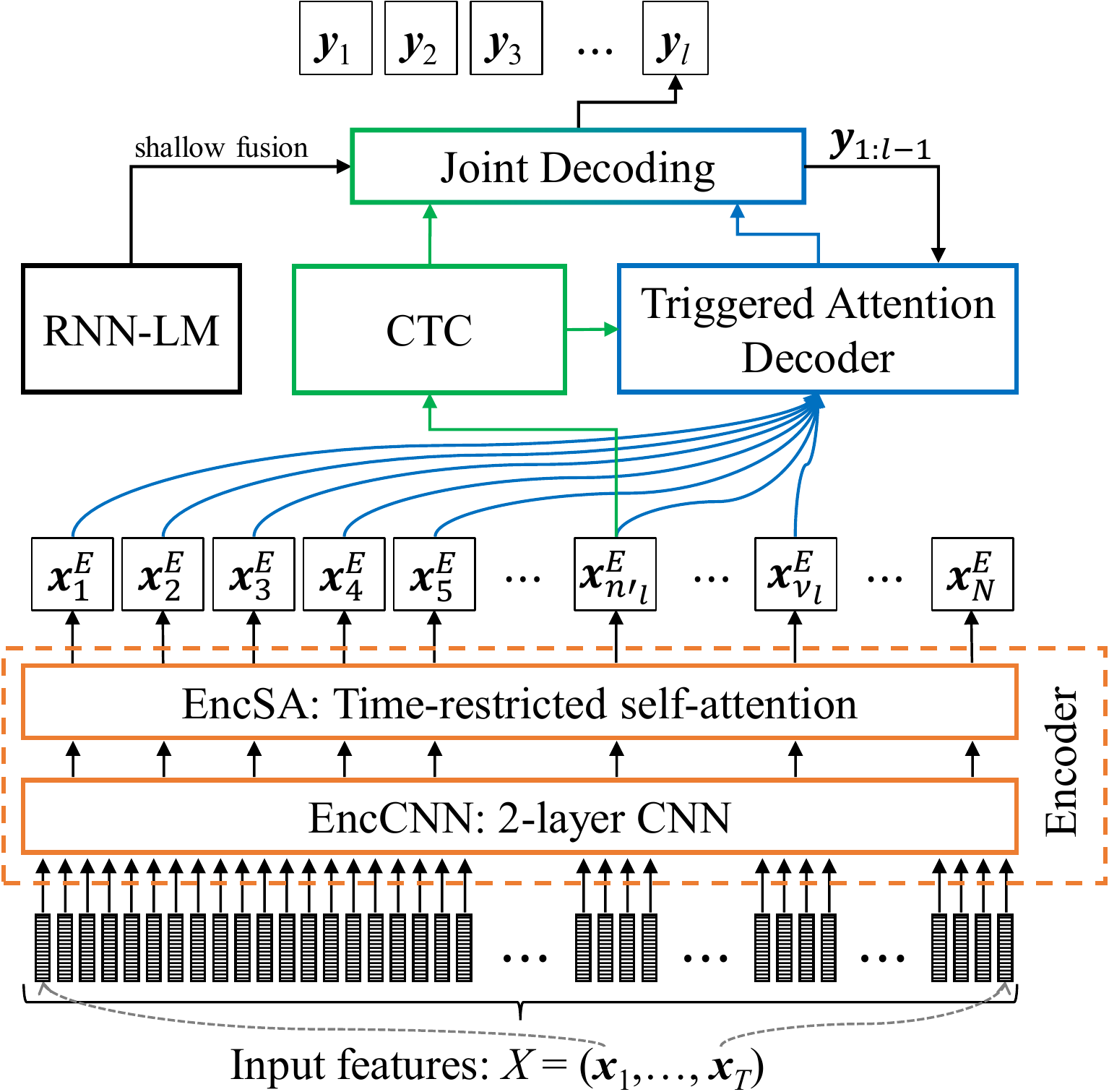}}
  \caption{Joint CTC-TA decoding scheme for streaming ASR with a transformer-based architecture.}
\label{fig:asr_system} %
\end{figure}

The streaming architecture of the proposed transformer-based ASR system is shown in Fig.~\ref{fig:asr_system}. The transformer is an encoder-decoder type of architecture that uses two different attention layers: encoder-decoder attention and self-attention.
The encoder-decoder attention can produce variable output lengths by using one or multiple query vectors, the decoder states, to control attention to a sequence of input values, the encoder state sequence.
In self-attention (SA), the queries, values, and keys are derived from the same input sequence, which results in an output sequence of the same length. Both attention types of the transformer model are based on the scaled dot-product attention mechanism,
\begin{equation}
    \text{Attention}(Q, K, V) = \text{Softmax}\left(\dfrac{Q K^\mathsf{T}}{\sqrt{d_k}}\right)V, \label{eq:mhatt} 
\end{equation}
where $Q \in \mathbb{R}^{n_q \times d_q}$, $K \in \mathbb{R}^{n_k \times d_k}$, and $V \in \mathbb{R}^{n_v \times d_v}$ are the queries, keys, and values, where the $d_*$ denote dimensions and the $n_*$ denote sequence lengths, $d_q=d_k$, and $n_k=n_v$ \cite{VaswaniSPUJGKP17}. Instead of using a single attention head, multiple attention heads are used by each layer of the transformer model with
\begin{align}
    \text{MHA}(\hat Q, \hat K, \hat V) &= \text{Concat}(\text{Head}_1, \dots, \text{Head}_{d_h}) W^\text{H} \label{eq:mha}\\
    \text{and } \text{Head}_i &= \text{Attention}(\hat Q W_i^Q, \hat K W_i^K, \hat V W_i^V), 
\end{align}
where $\hat Q$, $\hat K$, and $\hat V$ are inputs to the multi-head attention (MHA) layer, $\text{Head}_i$ represents the output of the $i$-th attention head for a total number of $d_h$ heads, and $W_i^Q \in \mathbb{R}^{d_\mathrm{model} \times d_q}$, $W_i^K \in \mathbb{R}^{d_\mathrm{model} \times d_k}$, $W_i^V \in \mathbb{R}^{d_\mathrm{model} \times d_v}$ as well as $W^H \in \mathbb{R}^{d_hd_v \times d_\mathrm{model}}$ are trainable weight matrices with typically $d_k=d_v=d_\mathrm{model}/d_h$.

\vspace{-0.1cm}
\subsection{Encoder: Time-restricted self-attention}
\label{ssec:encoder}
\vspace{-0.1cm}

The encoder of our transformer architecture consists of a two-layer CNN module \textsc{EncCNN} and a stack of $E$ self-attention layers \textsc{EncSA}:
\begin{align}
    X_0 &= \textsc{EncCNN} (X), \\
    X_{E} &= \textsc{EncSA} (X_0), \label{eq:enc_sa}
\end{align}
where $X=(\bm x_1, \dots, \bm x_T)$ denotes a sequence of acoustic input features, which are 80-dimensional log-mel spectral energies plus 3 extra features for pitch information \cite{HoriWZC17}. Both CNN layers of \textsc{EncCNN} use a stride of size $2$, a kernel size of $3 \times 3$, and a ReLU activation function. Thus, the striding reduces the frame rate of output sequence $X_0$ by a factor of 4 compared to the feature frame rate of $X$. 
The \textsc{EncSA} module of (\ref{eq:enc_sa}) consists of $E$ layers, where the $e$-th layer, for $e=1,\dots,E$, is a composite of a multi-head self-attention layer
\begin{equation}
    X'_e = X_{e-1} + \text{MHA}_{e}(X_{e-1}, X_{e-1}, X_{e-1}),
\end{equation}
and two feed-forward neural networks of inner dimension $d_\mathrm{ff}$ and outer dimension $d_\mathrm{model}$ that are separated by a ReLU activation function as follows:
\begin{align}
    X_{e} &= X'_e + \text{FF}_e(X'_e), \\
    \text{with } \text{FF}_e(X'_e) &= \text{ReLU} (X'_e W_{e,1}^\mathrm{ff} + b_{e,1}^\mathrm{ff}) W_{e,2}^\mathrm{ff} + b_{e,2}^\mathrm{ff},
\end{align}
where $W_{e,1}^\mathrm{ff} \in \mathbb{R}^{d_\mathrm{model} \times d_\mathrm{ff}}$, $W_{e,2}^\mathrm{ff} \in \mathbb{R}^{d_\mathrm{ff} \times d_\mathrm{model}}$, $b_{e,1}^\mathrm{ff} \in \mathbb{R}^{d_\mathrm{ff}}$, and $b_{e,2}^\mathrm{ff} \in \mathbb{R}^{d_\mathrm{model}}$ are trainable weight matrices and bias vectors. 

In order to control the latency of the encoder architecture, the future context of input sequence $X_0$ is limited to a fixed size, which is referred to as restricted or time-restricted self-attention \cite{VaswaniSPUJGKP17} and was first applied to hybrid HMM-based ASR systems \cite{PoveyHGLK18}. We can define a time-restricted self-attention encoder $\textsc{EncSA}^\mathrm{tr}$, with $n=1,\dots,N$, as
\begin{equation}
    \bm x_{1:n}^E = \textsc{EncSA}^\mathrm{tr} (\bm x_{1:n+\varepsilon^\text{enc}}^0), \label{eq:enc_sa_tr}
\end{equation}
where $\bm x_{1:n+\varepsilon^\text{enc}}^0 = X_0[1\mathbin{:}n+\varepsilon^\text{enc}] = (\bm x_1^0,\dots,\bm x_{n+\varepsilon^\text{enc}}^0)$, and $\varepsilon^\text{enc}$ denotes the number of look-ahead frames used by the time-restricted self-attention mechanism.

\vspace{-0.2cm}
\subsection{Decoder: Triggered attention}
\label{ssec:decoder}
\vspace{-0.2cm}

The encoder-decoder attention mechanism of the transformer model is using the TA concept \cite{MoritzHR19,MoritzHR19c} to enable the decoder to operate in a streaming fashion. TA training requires an alignment between the encoder state sequence $X_{E}$ and the label sequence $Y=(y_1, \dots, y_L)$ to condition the attention mechanism of the decoder only on past encoder frames plus a fixed number of look-ahead frames $\varepsilon^\text{dec}$. This information is generated by forced alignment using an auxiliary CTC objective $p_{\text{ctc}}(Y|X_E)$ \cite{GravesFGS06}, which is jointly trained with the decoder model, where the encoder neural network is shared \cite{KaritaCH19,WatanabeHKHH17,MoritzHR19}.

The triggered attention objective function is defined as
\begin{equation}
  p_{\text{ta}}(Y|X_E) = \prod_{l=1}^{L} p(y_l | \bm y_{1:l-1}, \bm x_{1:\nu_l}^E) \label{eq:trig_att}
\end{equation}
with $\nu_l = n'_l + \varepsilon^\text{dec}$, where $n'_l$ denotes the position of the first occurrence of label $y_l$ in the CTC forced alignment sequence \cite{MoritzHR19,MoritzHR19c}, $\bm y_{1:l-1}=(y_1,\dots,y_{l-1})$, and $\bm x_{1:\nu_l}^E=(\bm{x}_1^E,\dots,\bm{x}_{\nu_l}^E)$, which corresponds to the truncated encoder sequence. %
The term $p(y_l | \bm y_{1:l-1}, \bm x_{1:\nu_l}^E)$ represents the transformer decoder model %
\begin{equation}
    p(y_l | \bm y_{1:l-1}, \bm x_{1:\nu_l}^E) = \textsc{DecTA} (\bm x_{1:\nu_l}^E, \bm y_{1:l-1}),
\end{equation}
with
\begin{align}
  \bm z_{1:l}^{0} &= \textsc{Embed}(\langle \text{sos} \rangle, y_{1}, \dots, y_{l-1}), \\
  \overline{\bm z}_l^d &= \bm z_{l}^{d-1} + \text{MHA}_d^\mathrm{self}(\bm z_{l}^{d-1},\bm z_{1:l}^{d-1},\bm z_{1:l}^{d-1}), \\
  \overline{\overline{\bm z}}_l^d &= \overline{\bm z}_l^d + \text{MHA}_d^\mathrm{dec}(\overline{\bm z}_l^d,\bm x_{1:\nu_l}^E,\bm x_{1:\nu_l}^E), \\
  \bm z_l^d &= \overline{\overline{\bm z}}_l^d + \text{FF}_d(\overline{\overline{\bm z}}_l^d),
\end{align}
for $d=1,\dots,D$, where $D$ denotes the number of decoder layers. Function \textsc{Embed} converts the input label sequence $(\langle \text{s} \rangle, y_{1}, \dots, y_{l-1})$ into a sequence of trainable embedding vectors $\bm z_{1:l}^{0}$, where $\langle \text{sos} \rangle$ denotes the start of sentence symbol.
Function \textsc{DecTA} finally predicts the posterior probability of label $y_l$ by applying a fully-connected projection layer to $\bm z_{l}^{D}$ and a softmax distribution over that output.

The CTC model and the triggered attention model of (\ref{eq:trig_att}) are trained jointly using the multi-objective loss function
\begin{equation}
  \mathcal{L} = -\gamma \log p_\text{ctc} - (1-\gamma) \log p_\text{ta},
  \label{eq:loss}
\end{equation}
where hyperparameter $\gamma$ controls the weighting between the two objective functions $p_\text{ctc}$ and $p_\text{ta}$.

\vspace{-0.2cm}
\subsection{Positional encoding}
\label{ssec:pos_enc}
\vspace{-0.2cm}

Sinusoidal positional positional encodings ($\textsc{PE}$) are added to the sequences $X_0$ and $Z_0$, which can be written as
\begin{align}
    \textsc{PE}(\mathrm{pos},2i) &= \sin(\mathrm{pos}/10000^{2i/d_\mathrm{model}}), \\
    \textsc{PE}(\mathrm{pos},2i+1) &= \cos(\mathrm{pos}/10000^{2i/d_\mathrm{model}}),
\end{align}
where $\mathrm{pos}$ and $i$ are the position and dimension indices \cite{VaswaniSPUJGKP17}.

\subsection{Joint CTC-triggered attention decoding}
\label{ssec:joint_dec}

\begin{algorithm}[t]
\caption{Joint CTC-triggered attention decoding} \label{alg:decode}
\begin{algorithmic}[1]
\Procedure{Decode}{$X_E$, $p_\text{ctc}$, $\lambda$, $\alpha_0$, $\alpha$, $\beta$, $K$, $P$, $\theta_1$, $\theta_2$}
\State $\ell \gets (\langle\text{sos}\rangle,)$
\State $\Omega\gets \{\ell\}$, $\Omega_{\text{ta}}\gets \{\ell\}$
\State $p_{\text{nb}}(\ell)\gets 0$, $p_{\text{b}}(\ell)\gets 1$
\State $p_{\text{ta}}(\ell)\gets 1$
\For{$n=1,\dots,N$}
  \State $\Omega_{\text{ctc}}, p_{\text{nb}}, p_{\text{b}} \gets \textsc{ CTCPrefix}(p_\text{ctc}(n),\Omega, p_{\text{nb}}, p_{\text{b}})$
  \For{$\ell$ \textbf{in} $\Omega_{\text{ctc}}$} \Comment{Compute CTC prefix scores}
    \State $p_{\text{prfx}}(\ell) \gets p_{\text{nb}}(\ell) + p_{\text{b}}(\ell)$
    \State $\widehat{p}_{\text{prfx}}(\ell) \gets \log p_{\text{prfx}}(\ell) + \alpha_0 \log p_{\text{LM}}(\ell) + \beta |\ell|$
  \EndFor
  \State $\widehat{\Omega} \gets \textsc{Prune}(\Omega_\text{ctc}, \widehat{p}_{\text{prfx}}, K, \theta_1$)
  \For{$\ell \textbf{ in } \widehat{\Omega}$} \Comment{Delete old prefixes in $\Omega_{\text{ta}}$}
    \If{$\ell \textbf{ in } \Omega_{\text{ta}} \textbf{ and } \textsc{DCond}(\ell, \widehat{\Omega}, p_\text{ctc})$}
      \State delete $\ell$ in $\Omega_\text{ta}$
    \EndIf
  \EndFor
  \For{$\ell \textbf{ in } \widehat{\Omega}$} \Comment{Compute transformer scores}
    \If{$\ell \textbf{ not in } \Omega_{\text{ta}} \textbf{ and }\textsc{ACond}(\ell, \widehat{\Omega}, p_\text{ctc})$}
      \State $p_{\text{ta}}(\ell) \gets \textsc{DecTA}(\bm x_{1:n+\varepsilon^\text{dec}}^E,\ell)$
      \State add $\ell$ to $\Omega_{\text{ta}}$
    \EndIf
  \EndFor
  \For{$\ell \textbf{ in } \widehat{\Omega}$} \Comment{Compute joint scores}
    \State $\widehat{\ell}\gets \ell \textbf{ if }\ell \textbf{ in }\Omega_{\text{ta}} \textbf{ else } \ell_{:-1}$
    \State $p \gets \lambda \log p_{\text{prfx}}(\ell) + (1-\lambda)\log p_{\text{ta}}(\widehat{\ell})$
    \State $p_{\text{joint}}(\ell) \gets p + \alpha \log p_{\text{LM}}(\ell) + \beta |\ell|$
  \EndFor
  \State $\Omega \gets \textsc{Max}(\widehat{\Omega},p_{\text{joint}},P)$
  \State $\widehat{\Omega} \gets \textsc{Prune}(\widehat{\Omega}, \widehat{p}_{\text{prfx}}, P, \theta_2$)
  \State $\Omega \gets \Omega + \widehat{\Omega}$
  \State remove from $\Omega_\text{ta}$ prefixes rejected due to pruning
\EndFor
\State \textbf{return} $\textsc{Max}(\widehat{\Omega},p_{\text{joint}},1)$
\EndProcedure
\end{algorithmic}
\end{algorithm}

Algorithm \ref{alg:decode} shows the frame-synchronous one-pass decoding procedure for joint scoring of the CTC and transformer model outputs, which is similar to the decoding scheme described in \cite{MoritzHR19c}. The decoding algorithm is based on the frame-synchronous prefix beam search algorithm of \cite{MaasHJN14}, extending it by integrating the triggered attention decoder.
The joint hypothesis set $\Omega$ and the TA hypothesis set $\Omega_{\text{ta}}$ are initialized in line 3 with the prefix sequence $\ell=(\langle\text{sos}\rangle,)$, where the symbol $\langle\text{sos}\rangle$ denotes the start of sentence label.
The CTC prefix beam search algorithm of \cite{MaasHJN14} maintains two separate probabilities for a prefix ending in blank $p_{\text{b}}$ and not ending in blank $p_{\text{nb}}$, which are initialized in line 4. The initial TA scores $p_\text{ta}$ are defined in line 5.

The frame-by-frame processing of the CTC posterior probability sequence $p_\text{ctc}$ and the encoder state sequence $X_E$ is shown from line 5 to 26, where $p_\text{ctc}(n)$ denotes the CTC posterior probability distribution at frame $n$.
The function \textsc{CTCPrefix} follows the CTC prefix beam search algorithm described in \cite{MaasHJN14}, which extends the set of prefixes $\Omega$ using the CTC posterior probabilities $p_\text{ctc}$ of the current time step $n$ and returns the separate CTC prefix scores $p_\text{b}$ and $p_\text{nb}$ as well as the newly proposed set of prefixes $\Omega_\text{ctc}$. A local pruning threshold of 0.0001 is used by \textsc{CTCPrefix} to ignore labels of lower CTC probability. Note that no language model or any pruning technique is used by \textsc{CTCPrefix}, they will be incorporated in the following steps.

The prefix probabilities $p_\text{prfx}$ and scores $\widehat{p}_\text{prfx}$ are computed in lines 9 and 10, where $p_\text{LM}$ represents the language model (LM) probability and $|\ell|$ denotes the length of prefix sequence $\ell$ without counting the start of sentence label $\langle\text{sos}\rangle$.
The function $\textsc{Prune}$ prunes the set of CTC prefixes $\Omega_\text{ctc}$ in line 11 in two ways: first, the $K$ most probable prefixes are selected based on $\widehat p_\text{prfx}$, then every prefix of score smaller than $\max(\widehat p_\text{prfx}) - \theta_1$ is discarded, with $\theta_1$ being the beam width. The remaining set of prefixes is stored in $\widehat \Omega$.
From line 12 to 14, prefixes are removed from the set $\Omega_\text{ta}$ if they satisfy a delete condition $\textsc{DCond}$, and from line 15 to 18, TA scores are computed by function $\textsc{DecTA}$ if an add condition $\textsc{ACond}$ returns ``true''. The delete and add conditions are used to delete ``old'' TA scores computed at a non-optimal frame position and to delay the computation of TA scores, if a new CTC prefix appeared at a supposedly too early time frame.
The interested reader is referred to \cite{MoritzHR19c} for more details on both conditions. Note that our ASR experiments indicated that both conditions could be skipped without any WER degradation for the LibriSpeech task, which uses word-piece output labels, whereas their usage improves WERs for tasks like WSJ \cite{wsj1} with character-level label outputs.
Joint CTC-TA scores, computed from line 19 to 22, are used to select the $P$ most probable prefixes for further processing, which are stored in set $\Omega$ as shown in line 23. In line 24, the set of CTC prefixes $\widehat \Omega$ is further pruned to a maximum number of $P$ prefixes with prefix scores within the beam width $\theta_2$. Line 25 adds the CTC prefix set $\widehat \Omega$ to the best jointly scored prefix set $\Omega$, and line 26 removes prefixes from $\Omega_\text{ta}$ that are no longer in $\Omega$ for the current and previous time steps.
Finally, \textsc{Decode} returns the prefix sequence of highest joint probability $p_\text{joint}$ as shown in line 27.

\section{Experiments}
\label{sec:experiments}

\subsection{Dataset}
\label{ssec:data}

The LibriSpeech data set, which is a speech corpus of read English audio books \cite{PanayotovCPK15}, is used to benchmark ASR systems presented in this work. LibriSpeech is based on the open-source project LibriVox and provides about 960 hours of training data, 10.7 hours of development data, and 10.5 hours of test data, whereby the development and test data sets are both split into approximately two halves named ``clean'' and ``other''. The separation into clean and other is based on the quality of the recorded utterance, which was assessed using an ASR system \cite{PanayotovCPK15}. 

\subsection{Settings}
\label{ssec:settings}

Two transformer model sizes are used in this work: \textit{small} and \textit{large}. Parameter settings of the small transformer model are $d_\mathrm{model}=256$, $d_\mathrm{ff}=2048$, $d_h=4$, $E=12$, and $D=6$, whereas the large transformer model uses $d_\mathrm{model}=512$ and $d_h=8$ instead. The Adam optimizer with $\beta_1=0.9$, $\beta_2=0.98$, $\epsilon=10^{-9}$ and learning rate scheduling similar to \cite{VaswaniSPUJGKP17} is applied for training using 25000 warmup steps. The initial learning rate is set to 10.0 and the number of training epochs amounts to 100 for the small model and to 120 for the large model setup \cite{KaritaYWD19,DongXX18}. 
The set of label outputs consists of 5000 subwords obtained by the SentencePiece method \cite{KudoR18}. Weight factor $\gamma$, which is used to balance the CTC and transformer model objectives during training, is set to 0.3. Layer normalization is applied before and dropout with a rate of 10\% after each $\text{MHA}$ and $\text{FF}$ layer. In addition, label smoothing with a penalty of 0.1 is used \cite{ParkCZC19}.
An RNN-based language model (LM) is employed via shallow fusion. The RNN-LM consists of 4 LSTM layers with 2048 units each trained using stochastic gradient descent and the official LM training text data of LibriSpeech \cite{PanayotovCPK15}.

The LM weight, CTC weight, and beam size of the full-sequence based joint CTC-attention decoding method are set to 0.7, 0.5, and 20 for the small transformer model and to 0.6, 0.4, and 30 for the large model setup. The parameter settings for CTC prefix beam search decoding \cite{MaasHJN14} are LM weight $\alpha_0=0.7$, pruning beam width $\theta_1=16.0$, insertion bonus $\beta=2.0$, and pruning size $K=30$. Parameters for joint CTC-TA decoding are CTC weight $\lambda=0.5$, CTC LM weight $\alpha_0=0.7$, LM weight $\alpha=0.5$, pruning beam width $\theta_1=16.0$, pruning beam width $\theta_2=6.0$, insertion bonus $\beta=2.0$, pruning size $K=300$, and pruning size $P=30$. All decoding hyperparameter settings are determined using the development data sets of LibriSpeech.

\subsection{Results}
\label{ssec:results}

\begin{table}[tb]
 \begin{center}
  \caption{WERs [\%] of the full-sequence based CTC-transformer model. Results are shown for joint CTC-attention decoding \cite{WatanabeHKHH17}, CTC prefix beam search decoding only \cite{MaasHJN14}, and attention beam search decoding only \cite{KaritaYWD19}. In addition, results for including the RNN-LM, for using data augmentation \cite{ParkCZC19} as well as for the large transformer setup are shown.}
  \label{tab:bl_results}
  \sisetup{table-format=2.1,round-mode=places,round-precision=1,table-number-alignment = center,detect-weight=true,detect-inline-weight=math}
  \resizebox{\linewidth}{!}
  {\setlength{\tabcolsep}{2pt}\begin{tabular}{lcccc|cccc|cccc}%
  \toprule
  & \multicolumn{4}{c}{CTC-attention dec.} & \multicolumn{4}{c}{CTC beam search} & \multicolumn{4}{c}{Att. beam search} \\ \cmidrule(lr){2-5}\cmidrule(lr){6-9}\cmidrule(lr){10-13} %
  & \multicolumn{2}{c}{clean} & \multicolumn{2}{c}{other} & \multicolumn{2}{c}{clean} & \multicolumn{2}{c}{other} & \multicolumn{2}{c}{clean} & \multicolumn{2}{c}{other} \\ %
  \cmidrule(lr){2-3}\cmidrule(lr){4-5}\cmidrule(lr){6-7}\cmidrule(lr){8-9}\cmidrule(lr){10-11}\cmidrule(lr){12-13}
  System & {dev} & {test} & {dev} & {test} & {dev} & {test} & {dev} & {test} & {dev} & {test} & {dev} & {test}\\
  \midrule
  baseline     & 4.7 & 4.9 & 13.0 & 12.9 & 6.1 & 6.1 & 15.7 & 15.9 & 6.0 & 7.8 & 14.5 & 14.9 \\
  ~+RNN-LM     & 2.9 & 3.1 & \phantom{1}8.0 & \phantom{1}8.4 & 3.1 & 3.4 & \phantom{1}9.3 & \phantom{1}9.6 & 4.7 & 7.2 & 10.7 & 11.5 \\
  ~+SpecAug.  & 2.4 & 2.8 & \phantom{1}6.4 & \phantom{1}6.7 & 2.9 & 3.2 & \phantom{1}7.6 & \phantom{1}7.9 & 4.2 & 5.2 & \phantom{1}8.3 & \phantom{1}8.6 \\
  ~+large     & 2.4 & 2.7 & \phantom{1}6.0 & \phantom{1}6.1 & 2.5 & 2.8 & \phantom{1}6.9 & \phantom{1}7.0 & 4.1 & 5.0 & \phantom{1}7.9 & \phantom{1}8.0 \\
\bottomrule
  \end{tabular}}
  \end{center}
  \vspace{-3mm}
\end{table}

Table~\ref{tab:bl_results} presents ASR results of our transformer-based baseline systems, which are jointly trained with CTC to optimize training convergence and ASR accuracy \cite{KaritaYWD19,WatanabeHKHH17}. Results of different decoding methods are shown with and without using the RNN-LM, SpecAugment \cite{ParkCZC19}, and the large transformer model. Table~\ref{tab:bl_results} demonstrates that joint CTC-attention decoding provides significantly better ASR results compared to CTC or attention decoding alone, whereas CTC prefix beam search decoding attains lower WERs compared to attention beam search decoding, except for the dev-clean, dev-other, and test-other conditions when no LM is used.
For attention beam search decoding, we normalize the log posterior probabilities of the transformer model and the RNN-LM scores when combining both using the hypothesis lengths \cite{KaritaCH19}. Still our attention results are worse compared to the CTC results, which is unexpected but demonstrates that joint decoding stabilizes the transformer results.

Table~\ref{tab:ta_results} shows WERs of the full-sequence and the time-restricted self-attention encoder architectures combined with the CTC prefix beam search decoding method of \cite{MaasHJN14} and our joint CTC-TA decoding method of Section~\ref{ssec:joint_dec}, which are both algorithms for streaming recognition.
Different encoder look-ahead settings are compared using $\varepsilon^\mathrm{enc}=0,1,2,3$, and $\infty$, where each consumed frame of the self-attention encoder corresponds to 40~ms of input due to the output frame rate of \textsc{EncCNN}. Since such look-ahead is applied at every encoder layer ($E=12$), the theoretical latency caused by the time-restricted self-attention encoder amounts to $E \times \varepsilon^\mathrm{enc} \times 40~\text{ms}$, i.e., to 0~ms ($\varepsilon^\mathrm{enc}=0$), 
480~ms ($\varepsilon^\mathrm{enc}=1$), 960~ms ($\varepsilon^\mathrm{enc}=2$), and 1440~ms ($\varepsilon^\mathrm{enc}=3$), respectively.
The CTC prefix beam search decoding results of Table~\ref{tab:ta_results} show that increasing $\varepsilon^\mathrm{enc}$ significantly improves the ASR accuracy, e.g., test-other WER drops from 9.4\% to 7.0\% when moving from 0 to $\infty$ (full-sequence) encoder look-ahead frames.
The influence of different TA decoder settings are compared in Table~\ref{tab:ta_results} as well, using $\varepsilon^\mathrm{dec}=6,12$, and $18$ look-ahead frames. Note that unlike the encoder, the total decoder delay does not grow with its depth, since each decoder layer is attending to the encoder output sequence $X_E$. Thus, the TA decoder delay amounts to $\varepsilon^\mathrm{dec} \times 40~\text{ms}$, i.e., to 240~ms ($\varepsilon^\mathrm{dec}=6$), 480~ms ($\varepsilon^\mathrm{dec}=12$), and 720~ms ($\varepsilon^\mathrm{dec}=18$), respectively.
Results show that joint CTC-TA decoding consistently improves WERs compared to CTC prefix beam search decoding, while for larger look-ahead values WERs are approaching the full-sequence CTC-attention decoding results, which can be noticed by comparing results of the $\varepsilon^\mathrm{enc}=\infty$, $\varepsilon^\mathrm{dec}=18$ TA system setup with the full-sequence CTC-attention system of Table~\ref{tab:bl_results}.

\begin{table}[tb]
 \begin{center}
  \caption{WERs [\%] for different $\varepsilon^\mathrm{enc}$ settings of the time-restricted encoder using the CTC prefix beam search decoding method of \cite{MaasHJN14} as well our proposed joint CTC-TA decoding method of Section~\ref{ssec:joint_dec} with different $\varepsilon^\mathrm{dec}$ configurations. %
  SpecAugment \cite{ParkCZC19}, the RNN-LM, and the large transformer are applied for all systems.\protect\footnotemark[1]}
  \label{tab:ta_results}
  \vspace{0.05cm}
  \resizebox{\linewidth}{!}
{\setlength{\tabcolsep}{2pt}\begin{tabular}{lcccc|cccc|cccc|cccc}
\toprule
  & \multicolumn{4}{c}{CTC beam search} & \multicolumn{4}{c}{TA: $\varepsilon^\mathrm{dec}=6$} & \multicolumn{4}{c}{TA: $\varepsilon^\mathrm{dec}=12$} & \multicolumn{4}{c}{TA: $\varepsilon^\mathrm{dec}=18$} \\ 
  \cmidrule(lr){2-5}\cmidrule(lr){6-9}\cmidrule(lr){10-13}\cmidrule(lr){14-17} %
  & \multicolumn{2}{c}{clean} & \multicolumn{2}{c}{other} & \multicolumn{2}{c}{clean} & \multicolumn{2}{c}{other} & \multicolumn{2}{c}{clean} & \multicolumn{2}{c}{other} & \multicolumn{2}{c}{clean} & \multicolumn{2}{c}{other} \\
  \cmidrule(lr){2-3}\cmidrule(lr){4-5}\cmidrule(lr){6-7}\cmidrule(lr){8-9}\cmidrule(lr){10-11}\cmidrule(lr){12-13}\cmidrule(lr){14-15}\cmidrule(lr){16-17}
  $\varepsilon^\mathrm{enc}$ & dev & test & dev & test & dev & test & dev & test & dev & test & dev & test & dev & test & dev & test \\
  \midrule
  $0$      & 3.3 & 3.7 & 9.4 & 9.4  & 3.2 & 3.3 & 8.4 & 8.6  & 3.0 & 3.4 & 8.4 & 8.5  & 2.9 & 3.2 & 8.1 & 8.0 \\ 
  $1$      & 3.0 & 3.3 & 8.4 & 8.6  & 2.9 & 3.1 & 7.8 & 8.1  & 2.8 & 3.1 & 7.5 & 8.1  & 2.8 & 3.0 & 7.5 & 7.8 \\
  $2$      & 2.9 & 3.1 & 8.0 & 8.2  & 2.8 & 2.9 & 7.4 & 7.8  & 2.7 & 2.9 & 7.2 & 7.6  & 2.7 & 2.9 & 7.3 & 7.4 \\
  $3$      & 2.8 & 2.9 & 7.8 & 8.1  & 2.7 & 2.8 & 7.2 & 7.4  & 2.7 & 2.8 & 7.2 & 7.3  & 2.7 & 2.8 & 7.1 & 7.2  \\
  $\infty$ & 2.5 & 2.8 & 6.9 & 7.0  & 2.5 & 2.7 & 6.3 & 6.5  & 2.5 & 2.7 & 6.3 & 6.4  & 2.4 & 2.6 & 6.1 & 6.3 \\
  \bottomrule
  \end{tabular}}
  \end{center}
  \vspace{-3mm}
\end{table}
\footnotetext[1]{Note that results shown here are updated compared to our ICASSP submission.}

The best streaming ASR system of Table~\ref{tab:ta_results} achieves a WER of 2.8\% and 7.2\% for the test-clean and test-other conditions of LibriSpeech with an overall processing delay of 30~ms (\textsc{EncCNN}) + 1440~ms (\textsc{EncSA}: $\varepsilon^\mathrm{enc}=3$) + 720~ms (\textsc{DecTA}: $\varepsilon^\mathrm{dec}=18$) = 2190~ms. For $\varepsilon^\mathrm{enc}=1$ and $\varepsilon^\mathrm{dec}=18$, the test-clean and test-other WERs amount to 3.0\% and 7.8\%, respectively,  with a total delay of 1230~ms, which provides a good trade-off between accuracy and latency.
It should be noted that a lattice-based CTC-TA decoding implementation can output intermediate CTC prefix beam search results, which are updated after joint scoring with the TA decoder, and thus the perceived latency of such an implementation will be on average smaller than its theoretical latency and close to that of the encoder alone. However, a thorough study of the user perceived latency remains to be done in future work.

\section{Conclusions}
\label{sec:conclusions}

In this paper, a fully streaming end-to-end ASR system based on the transformer architecture is proposed. Time-restricted self-attention is applied to control the latency of the encoder and the triggered attention (TA) concept to control the output latency of the decoder. For streaming recognition and joint CTC-transformer model scoring, a frame-synchronous one-pass decoding algorithm is applied, which demonstrated similar LibriSpeech ASR results compared to full-sequence based CTC-attention as the number of look-ahead frames is increased. Combined with the time-restricted self-attention encoder, our proposed TA-based streaming ASR system achieved WERs of $2.8\%$ and $7.2\%$ for the test-clean and test-other data sets of LibriSpeech, which to our knowledge is the best published LibriSpeech result of a fully streaming end-to-end ASR system.

\vfill\pagebreak

\bibliographystyle{IEEEtran}
\bibliography{refs}

\begin{thebibliography}{10}
\providecommand{\url}[1]{#1}
\csname url@samestyle\endcsname
\providecommand{\newblock}{\relax}
\providecommand{\bibinfo}[2]{#2}
\providecommand{\BIBentrySTDinterwordspacing}{\spaceskip=0pt\relax}
\providecommand{\BIBentryALTinterwordstretchfactor}{4}
\providecommand{\BIBentryALTinterwordspacing}{\spaceskip=\fontdimen2\font plus
\BIBentryALTinterwordstretchfactor\fontdimen3\font minus
  \fontdimen4\font\relax}
\providecommand{\BIBforeignlanguage}[2]{{%
\expandafter\ifx\csname l@#1\endcsname\relax
\typeout{** WARNING: IEEEtran.bst: No hyphenation pattern has been}%
\typeout{** loaded for the language `#1'. Using the pattern for}%
\typeout{** the default language instead.}%
\else
\language=\csname l@#1\endcsname
\fi
#2}}
\providecommand{\BIBdecl}{\relax}
\BIBdecl

\bibitem{PoveyPGG16}
D.~Povey, V.~Peddinti, D.~Galvez, P.~Ghahremani, V.~Manohar, X.~Na, Y.~Wang,
  and S.~Khudanpur, ``Purely sequence-trained neural networks for {ASR} based
  on lattice-free {MMI},'' in \emph{Proc. ISCA Interspeech}, Sep. 2016, pp.
  2751--2755.

\bibitem{HintonDYD12}
G.~{Hinton}, L.~{Deng}, D.~{Yu}, G.~E. {Dahl}, A.~{Mohamed}, N.~{Jaitly},
  A.~{Senior}, V.~{Vanhoucke}, P.~{Nguyen}, T.~N. {Sainath}, and
  B.~{Kingsbury}, ``Deep neural networks for acoustic modeling in speech
  recognition: The shared views of four research groups,'' \emph{IEEE Signal
  Processing Magazine}, vol.~29, no.~6, pp. 82--97, 2012.

\bibitem{KaritaYWD19}
S.~Karita, N.~Yalta, S.~Watanabe, M.~Delcroix, A.~Ogawa, and T.~Nakatani,
  ``Improving transformer-based end-to-end speech recognition with
  connectionist temporal classification and language model integration,'' in
  \emph{Proc. ISCA Interspeech}, Sep. 2019, pp. 1408--1412.

\bibitem{GravesFGS06}
A.~Graves, S.~Fern{\'{a}}ndez, F.~J. Gomez, and J.~Schmidhuber, ``Connectionist
  temporal classification: labelling unsegmented sequence data with recurrent
  neural networks,'' in \emph{Proc. International Conference on Machine
  Learning (ICML)}, vol. 148, Jun. 2006, pp. 369--376.

\bibitem{Graves12}
A.~Graves, ``Sequence transduction with recurrent neural networks,''
  \emph{arXiv preprint arXiv:abs/1211.3711}, 2012.

\bibitem{BahdanauCB14}
D.~Bahdanau, K.~Cho, and Y.~Bengio, ``Neural machine translation by jointly
  learning to align and translate,'' \emph{arXiv preprint arXiv:abs/1409.0473},
  2014.

\bibitem{Schalkwyk19}
J.~Schalkwyk, ``An all-neural on-device speech recognizer,'' Mar. 2019, url:
  https://ai.googleblog.com/2019/03/an-all-neural-on-device-speech.html.

\bibitem{LiZHG19}
J.~Li, R.~Zhao, H.~Hu, and Y.~Gong, ``Improving {RNN} transducer modeling for
  end-to-end speech recognition,'' \emph{arXiv preprint arXiv:abs/1909.12415},
  2019.

\bibitem{PrabhavalkarRSL17}
R.~Prabhavalkar, K.~Rao, T.~N. Sainath, B.~Li, L.~Johnson, and N.~Jaitly, ``A
  comparison of sequence-to-sequence models for speech recognition,'' in
  \emph{Proc. ISCA Interspeech}, Sep. 2017, pp. 939--943.

\bibitem{SainathCPKWNC17}
T.~N. Sainath, C.~Chiu, R.~Prabhavalkar, A.~Kannan, Y.~Wu, P.~Nguyen, and
  Z.~Chen, ``Improving the performance of online neural transducer models,''
  \emph{arXiv preprint arXiv:abs/1712.01807}, 2017.

\bibitem{ChiuR18}
C.~Chiu and C.~Raffel, ``Monotonic chunkwise attention,'' in \emph{Proc.
  International Conference on Learning Representations (ICLR)}, Apr. 2018.

\bibitem{MoritzHR19}
N.~{Moritz}, T.~{Hori}, and J.~{Le Roux}, ``Triggered attention for end-to-end
  speech recognition,'' in \emph{Proc. IEEE International Conference on
  Acoustics, Speech and Signal Processing (ICASSP)}, May 2019, pp. 5666--5670.

\bibitem{WatanabeHKHH17}
S.~Watanabe, T.~Hori, S.~Kim, J.~R. Hershey, and T.~Hayashi, ``Hybrid
  {CTC}/attention architecture for end-to-end speech recognition,'' \emph{J.
  Sel. Topics Signal Processing}, vol.~11, no.~8, pp. 1240--1253, 2017.

\bibitem{MoritzHR19c}
N.~Moritz, T.~Hori, and J.~{Le Roux}, ``Streaming end-to-end speech recognition
  with joint {CTC}-attention based models,'' in \emph{Proc. IEEE Workshop on
  Automatic Speech Recognition and Understanding (ASRU)}, Dec. 2019, pp.
  936--943.

\bibitem{MoritzHR19b}
N.~{Moritz}, T.~{Hori}, and J.~{Le Roux}, ``Unidirectional neural network
  architectures for end-to-end automatic speech recognition,'' in \emph{Proc.
  ISCA Interspeech}, Sep. 2019, pp. 76--80.

\bibitem{VaswaniSPUJGKP17}
A.~Vaswani, N.~Shazeer, N.~Parmar, J.~Uszkoreit, L.~Jones, A.~N. Gomez,
  L.~Kaiser, and I.~Polosukhin, ``Attention is all you need,'' in \emph{Proc.
  NIPS}, Dec. 2017, pp. 6000--6010.

\bibitem{KaritaCH19}
S.~Karita, N.~Chen, T.~Hayashi, T.~Hori, H.~Inaguma, Z.~Jiang, M.~Someki,
  N.~E.~Y. Soplin, R.~Yamamoto, X.~Wang, S.~Watanabe, T.~Yoshimura, and
  W.~Zhang, ``A comparative study on transformer vs {RNN} in speech
  applications,'' in \emph{Proc. IEEE Workshop on Automatic Speech Recognition
  and Understanding (ASRU)}, Dec. 2019.

\bibitem{HoriWZC17}
T.~Hori, S.~Watanabe, Y.~Zhang, and W.~Chan, ``Advances in joint
  {CTC}-attention based end-to-end speech recognition with a deep {CNN} encoder
  and {RNN-LM},'' in \emph{Proc. ISCA Interspeech}, Aug. 2017, pp. 949--953.

\bibitem{PoveyHGLK18}
D.~{Povey}, H.~{Hadian}, P.~{Ghahremani}, K.~{Li}, and S.~{Khudanpur}, ``A
  time-restricted self-attention layer for {ASR},'' in \emph{Proc. IEEE
  International Conference on Acoustics, Speech and Signal Processing
  (ICASSP)}, 2018, pp. 5874--5878.

\bibitem{MaasHJN14}
A.~L. Maas, A.~Y. Hannun, D.~Jurafsky, and A.~Y. Ng, ``First-pass large
  vocabulary continuous speech recognition using bi-directional recurrent
  {DNN}s,'' \emph{arXiv preprint arXiv:1408.2873}, 2014.

\bibitem{wsj1}
``{CSR}-{II} ({WSJ1}) complete,'' vol. LDC94S13A.\hskip 1em plus 0.5em minus
  0.4em\relax Philadelphia: Linguistic Data Consortium, 1994.

\bibitem{PanayotovCPK15}
V.~Panayotov, G.~Chen, D.~Povey, and S.~Khudanpur, ``Librispeech: An {ASR}
  corpus based on public domain audio books,'' in \emph{Proc. IEEE
  International Conference on Acoustics, Speech and Signal Processing
  (ICASSP)}, Apr. 2015, pp. 5206--5210.

\bibitem{DongXX18}
L.~Dong, S.~Xu, and B.~Xu, ``Speech-transformer: {A} no-recurrence
  sequence-to-sequence model for speech recognition,'' in \emph{Proc. IEEE
  International Conference on Acoustics, Speech and Signal Processing
  (ICASSP)}, 2018, pp. 5884--5888.

\bibitem{KudoR18}
T.~Kudo and J.~Richardson, ``Sentence{P}iece: {A} simple and language
  independent subword tokenizer and detokenizer for neural text processing,''
  \emph{arXiv preprint arXiv:abs/1808.06226}, 2018.

\bibitem{ParkCZC19}
D.~S. Park, W.~Chan, Y.~Zhang, C.-C. Chiu, B.~Zoph, E.~D. Cubuk, and Q.~V. Le,
  ``{S}pec{A}ugment: {A} simple data augmentation method for automatic speech
  recognition,'' \emph{arXiv preprint arXiv:abs/1904.08779}, 2019.

\end{thebibliography}

\end{document}